\documentclass[11pt]{article}

\usepackage{booktabs}  % tables
\usepackage{siunitx}
\usepackage{graphicx}
\usepackage{wrapfig}
\usepackage{rotating}

\usepackage{amsmath}
\usepackage{amsthm}
\newcommand\rurl[1]{%
  \href{https://#1}{\nolinkurl{#1}}%
}

\usepackage[final]{automl}
\usepackage{natbib}
\title{The Potential of AutoML for Recommender Systems}
\author[1]{\nameemail{Tobias Vente}{tobias.vente@uni-siegen.de}}
\author[1]{\nameemail{Joeran Beel}{joeran.beel@uni-siegen.de}}

\affil[1]{University of Siegen}
\hypersetup{%
  pdfauthor={}, % will be reset to "Anonymous" unless the "final" package option is given
  pdftitle={},
  pdfsubject={},
  pdfkeywords={}
}

\begin{document}

\maketitle

\begin{abstract}
% Describe: Background, Problem, Goal, Methodology, Results education 
% Background
Automated Machine Learning (AutoML) has greatly advanced applications of Machine Learning (ML) including model compression, machine translation, and computer vision.
Recommender Systems (RecSys) can be seen as an application of ML. Yet, AutoML has found little attention in the RecSys community; nor has RecSys found notable attention in the AutoML community.
Only few and relatively simple Automated Recommender Systems (AutoRecSys) libraries exist that adopt AutoML techniques.
However, these libraries are based on student projects and do not offer the features and thorough development of AutoML libraries. 
% Goal:
We set out to determine how AutoML libraries perform in the scenario of an inexperienced user who wants to implement a recommender system.
% Methodology:
We compared the predictive performance of $60$ AutoML, AutoRecSys, ML, and RecSys algorithms from $15$ libraries, including a mean predictor baseline, on $14$ explicit feedback RecSys datasets.
% What is an inexperienced user 
To simulate the perspective of an inexperienced user, the algorithms were evaluated with default hyperparameters. 
% Results 
We found that AutoML and AutoRecSys libraries performed best. AutoML libraries performed best for six of the $14$ datasets ($43\%$), but it was not always the same AutoML library performing best. 
The single-best library was the AutoRecSys library Auto-Surprise, which performed best on five datasets ($36\%$).
On three datasets ($21\%$), AutoML libraries performed poorly, and RecSys libraries with default parameters performed best.
Although, while obtaining $50\%$ of all placements in the top five per dataset, RecSys algorithms fall behind AutoML on average. ML algorithms generally performed the worst. 
\end{abstract}
\section{Introduction}
%%% Should contain: Background, Motivation, Research Problem, Research question, Research Goal, Contribution
%% Background 
% Intro about accessible 
The application of Machine Learning (ML) was made more accessible by Automated Machine Learning (AutoML) \citep{DBLP:books/sp/HKV2019}. AutoML enables users with little ML knowledge to apply ML algorithms relatively easily and effectively in various fields. 
Sometimes, AutoML libraries may achieve even better performance than human experts with ML libraries \citep{hanussek2020can}. 

% What is Recsys
ML is often used to implement Recommender Systems (RecSys) \citep{DBLP:journals/csur/ZhangYST19}. The goal of RecSys is, among others, to predict the rating that a customer would give to an item (e.g., a movie or e-commerce product); or to predict the preference of a customer for an item ("like" / "dislike"). 
Implementing RecSys with ML or non-ML algorithms is as challenging as implementing any other ML pipeline.
Many different algorithms with many hyperparameters exist.
A user of RecSys algorithms must select algorithms and features, tune hyperparameters, and conduct time-consuming evaluations. If the user is inexperienced and implements the pipeline sub-optimally, then the recommendations wont satisfy customers and harm the business's success.

The success of AutoML has resulted in the development of a few Automated Recommender Systems (AutoRecSys) libraries that adopt AutoML techniques to RecSys \citep{anand2020auto,wang2020autorec,DBLP:conf/cikm/SonboliMGKLLSB21,Gupta2020}. However, these AutoRecSys libraries are based on student projects and not developed as professionally as many AutoML libraries.  
A user could also use AutoML by formulating a RecSys task as a ML classification or regression task. 

%% Problem
% Inexperienced user relationship  
Using ML or RecSys libraries is challenging for inexperienced users. %like developers or researchers
To satisfy customers, inexperienced users would need to find a suitable configuration. 
% Automated approaches
To do so, AutoML or AutoRecSys could help inexperienced users and make RecSys more accessible. 
AutoML for RecSys could have potential because no holistic and easily usable (Auto)RecSys libraries exist. This is in stark contrast to the plenty holistic and easily usable AutoML libraries.

% Detail problems of (Auto)RecSys libs
Many RecSys libraries exist. RecSys libraries are often designed for particular RecSys applications (e.g., only explicit feedback); built with a focus on specific types of algorithms (e.g., neural networks); and, depending on its community, actively maintained or not maintained at all. 
State-of-the-art algorithms could be too complicated to reproduce for an inexperienced user or not reproducible at all, see \citep{dacrema2019we}.
Using AutoRecSys can be problematic since users' needs, like implicit feedback, are not supported yet.   

%% Goal / Research Question 
It is an open question whether an inexperienced user can achieve suitable performance with AutoML compared to AutoRecSys, ML, or RecSys algorithms.
% Hence use default values 
We understand an inexperienced user to be a user that only employs default hyperparameters. 
% Explicit Goal / RQ
This motivates our research question: \emph{What is the performance of AutoML algorithms on RecSys tasks compared to AutoRecSys, ML, and RecSys algorithms using default hyperparameters?}

In this study, we do not explore the reasons for the differences in performance between algorithms. So far, nobody investigated the potential of AutoML for RecSys. Hence, we are first interested in an assessment of the state-of-the-art for inexperienced users as an initial exploratory study. A detailed analysis of the performance differences is too extensive for this paper and left for future work.

%% Methodology 
To answer our research question, we compare $7$ AutoML-, $28$ ML-, $23$ RecSys algorithms, $1$ AutoRecSys algorithm, and a mean predictor baseline on $14$ RecSys datasets. 
We used default values for all hyperparameters to simulate the perspective of an inexperienced user. 
We limit this comparison to explicit feedback, a RecSys rating prediction task, and offline evaluation.

%% Research Goal - vague goal for long-term oriented research
We hope that this work directs the interest of the AutoML community towards RecSys. 
Moreover, we hope that this work shows the special application of automated approaches in RecSys and the need thereof. 
%% Contribution - explicitly state the contribution of your work
We contribute insights into the potential of AutoML for an applied field like RecSys.
Furthermore, we collected and preprocessed $14$ datasets for predicting explicit feedback. 
Code for the comparison and for preprocessing the datasets can be found in our GitHub repository\footnote{\url{https://github.com/ISG-Siegen/AutoML_for_Recommender_Systems}}. 

\section{Related Work}
%%% Only provide information relevant to the research problem (Provide new meta information, Not just summarize but add new information, Critical but fair)
%% AutoML for RecSys
To the best of our knowledge, no one has evaluated AutoML libraries on a relatively large number of RecSys datasets. 
% AtuoML for Esenmbles: https://link.springer.com/article/10.1007/s13369-021-05670-z
The work closest to ours is the usage of AutoML to generate ensembles for RecSys \citep{gupta2021enpso}. 
% Apply RecSys to AUtoML: https://dl.acm.org/doi/abs/10.1145/3292500.3330909
RecSys algorithms were used for AutoML in, for example, model selection \citep{DBLP:conf/kdd/YangAKU19}.
%% Applications of AutoML to other applied fields
AutoML has been applied to other fields like
% computer vision: Medical Image Analysis: https://ieeexplore.ieee.org/document/9434062; Image Sentinal Analysis https://ieeexplore.ieee.org/document/9533552; disparity estimation https://openaccess.thecvf.com/content_ICCV_2019/html/Saikia_AutoDispNet_Improving_Disparity_Estimation_With_AutoML_ICCV_2019_paper.html
computer vision \citep{DBLP:conf/isbi/YangSN21, DBLP:conf/iccv/SaikiaMZHB19, DBLP:conf/ijcnn/LopesGAC21}, 
% Machine Translation: https://link.springer.com/chapter/10.1007/978-3-030-46140-9_29
machine translation \citep{DBLP:conf/simbig/ViswanathanWK19},
% compression of models: https://openaccess.thecvf.com/content_ECCV_2018/html/Yihui_He_AMC_Automated_Model_ECCV_2018_paper.html
model compression \citep{DBLP:conf/eccv/HeLLWLH18}, 
%Application of AutoML in the Automated Coding of Educational Discourse Data https://repository.isls.org/handle/1/6628;  educational data mining: https://www.mdpi.com/2076-3417/10/1/90
and education \citep{tsiakmaki2020implementing, DBLP:conf/icls/LeeG020}.
%% AutoML itself has already 
To compare different AutoML frameworks, an AutoML benchmark was created \citep{2_gijsbers2019open}. 
Moreover, in the AutoML community, large dataset collections exist \citep{bischl2017openml,balaji2018benchmarking}. 
For example, benchmarking suites\footnote{\url{https://docs.openml.org/benchmark/}} on OpenML \citep{OpenML2013}.
No such dataset collections exist in the recommender system community. 
Furthermore, there is no standardized benchmark framework to compare RecSys algorithms. 

\section{Methods}
\label{sec:benchmarkDesign}
%%% 1) Explain what and how we did it shortly 2) Justify relevant decisions
% What  
We evaluated the predictive performance of $7$ AutoML-, $28$ ML-, $23$ RecSys algorithms, $1$ AutoRecSys algorithm, and a mean predictor baseline on $14$ RecSys datasets. 
% How  
Each RecSys dataset contains explicit feedback, which represents a rating prediction task.
Our implementation made the datasets usable by (Auto)RecSys and (Auto)ML algorithms. 
To capture the predictive performance, we computed the Root Mean Squared Error (RMSE) using hold-out validation (random test split: $25\%$).
For our inexperienced-user-scenario, the algorithms are evaluated with default hyperparameters. 
To compare the $60$ algorithms, we created a Docker-based evaluation tool. 
Moreover, we created preprocessing scripts for all $14$ RecSys datasets.  
% Compute stats 
All $840$ evaluation runs ($60$ algorithms times $14$ datasets) were done on a workstation with an AMD Ryzen Threadripper PRO 3975WX CPU, 32 Cores (64 threads), SSD storage, and 528 GB RAM. 
The evaluations were run sequentially to guarantee that each run can utilize all resources. In total, it took 25 days to run all evaluations. 

%% Why not AutoML Benchmark reused 
We did not reuse the AutoML benchmark \citep{2_gijsbers2019open} for two reasons. 
% regression
First, the benchmark's support for regression is a work in progress, which is, however, required for the task of predicting ratings.
% resource limits
Second, the benchmark does not fully control runtime and memory limits independently from the evaluated frameworks. It assumes that the frameworks constrain the resources themselves. 
For example, the maximum runtime for an evaluation is passed to the framework and only if this constraint is not respected after twice the maximum runtime, the evaluation is aborted\footnote{\url{https://github.com/openml/automlbenchmark/blob/master/docs/HOWTO.md}}. Moreover, it is not a requirement that the frameworks limit memory. 
% An alternative would be to use the benchmark's profiling tool to determine resource limitation discrepancies. 
% not to automl benchmark
For our use case of regression and our need of resource limits due to budget constraints, the AutoML benchmark was not appropriate. 

%% Why only preprocessing scripts  
All $14$ RecSys datasets are publicly available. However, we can not redistribute the datasets due to their licenses. 
We only provide our preprocessing scripts so that others can replicate the evaluated datasets.
% Difference RecSys to ML dataset
RecSys datasets for explicit feedback differ from normal ML datasets in that RecSys datasets contain customer-, item-IDs, and timestamps as features.
The timestamp, or a feature that can be transformed into a timestamp (e.g.: date of rating), represents the point in time a customer rated an item. 
Additional features are optional and not necessarily used by RecSys algorithms.  
The $23$ RecSys algorithms included in our comparison only use customer-, item-IDs, and optionally timestamps. 
Consequently, we compare (Auto)RecSys using two or three features with (Auto)ML algorithms using all features. 
%The labels of datasets are ratings, numeric values with bounds. 
RecSys datasets are not synthetic but derived from real-world applications including human-derived data.
% Overview of upcoming subsections 
In the following, we first present the selected datasets. Then, we detail the compared libraries. Lastly, we talk about our experiment setup by detailing the preprocessing, hyperparameters, and resource limits. 

\subsection{Datasets}
\label{sec:Datasets}
%% Overview
We selected the following $14$ RecSys datasets for explicit feedback\footnote{The URLs to and licenses of all datasets can be found in the appendix \ref{apdx:lib_data}.}. 
These datasets are some of the most common and public rating-related RecSys tasks, which we found through the dataset survey by \cite{beel2019data} and papers from RecSys conferences. 
\begin{itemize}
% Movielens % +3 (Movielens) 
\item \textbf{MovieLens Data:} We have used the MovieLens 100k,
MovieLens 1M, and MovieLens Latest 100k (9/2018) datasets \citep{DBLP:journals/tiis/HarperK16}.

% Amazon Review % + 8 (Amazon)
\item \textbf{Amazon Review Data:} The datasets based on the following categories were used: Electronics; Digital Music; Toys and Games; Movies and TV; Fashion; Appliances; Industrial and Scientific; Software. We used the 5-core versions combined with meta-data \citep{ni2019justifying}.

% +1 (Yelp) % +1 (Netflix) % + 1 (Food.com)
\item \textbf{Other:} We have also used the Yelp Open Dataset \citep{yelpdataset}; the Netflix Prize Dataset \citep{bennett2007netflix}\footnote{We randomly subsampled this dataset to ten million instances due to our budget constraints.}; and the Food.com Recipes and Interactions Dataset \citep{majumder2019generating}.
\end{itemize}

\subsection{Libraries}
\label{sec:libraries}
%We used the following libraries for our comparison. 
All selected libraries are open source and can predict explicit feedback\footnote{The versions, URLs to, and licenses of all libraries can be found in the appendix \ref{apdx:lib_data}.}.
\begin{itemize}
% RecSys
\item \textbf{RecSys:} 
Surprise \citep{Hug2020}, 
LensKit \citep{ekstrand2020lenskit}, 
Spotlight \citep{kula2017spotlight}, 
AutoRec \citep{wang2020autorec} 

% AutoRecSys
\item \textbf{AutoRecSys:}
Auto-Surprise \citep{anand2020auto}

% ML
\item \textbf{ML:}  
scikit-learn \citep{scikit-learn}, 
XGBoost \citep{chen2016xgboost}, 
ktrain \citep{maiya2020ktrain}

% AutoML
\item \textbf{AutoML:}
Auto-sklearn \citep{feurer-neurips15a}, 
FLAML \citep{wang2021flaml}, 
GAMA \citep{Gijsbers2019}, 
H2O \citep{H2OAutoML20}, 
TPOT \citep{OlsonGECCO2016,le2020scaling}, 
AutoGluon \citep{agtabular}, 
Auto-PyTorch \citep{zimmer-tpami21a, mendoza-automlbook18a}
\end{itemize}

%% Provide some Information about Auto-Surprise
The Auto-Surprise \citep{anand2020auto} library is an extension of the Surprise \citep{Hug2020} library. Auto-Surprise uses the Tree of Parzen Estimators of Hyperopt \citep{Bergstra_2015} to solve the Combined Algorithm and Hyper-parameter Selection (CASH) problem for a search space created from the algorithms and hyperparameter of Surprise. Auto-Surprise is currently the only library that we labeled as AutoRecSys because it is automated and sophisticated enough. 

%% Discuss the problem of AutoRec, LibRec, AutoCaseRec
% AutoRec
We labeled the library AutoRec as a RecSys library and not as an AutoRecSys library. 
AutoRec provides an interface to use AutoKeras \citep{jin2019auto, chollet2015keras} without automating the additional overhead of the interface itself.  
For example, AutoRec does not enable the usage of a time limit for the neural architecture search nor does it automatically setup the input interaction matrix.
Hence, we deem AutoRec to be not automated enough to be called AutoRecSys. 
% Other tools
Besides Auto-Surprise and AutoRec, two other projects related to AutoRecSys exist: Librec-Auto\footnote{\url{https://github.com/that-recsys-lab/librec-auto}} \citep{DBLP:conf/cikm/SonboliMGKLLSB21} and Auto-CaseRec\footnote{\url{https://github.com/BeelGroup/Auto-CaseRec}} \citep{Gupta2020}.

% Librec-Auto
Librec-Auto takes as input one or more algorithms and their configurations defined in a XML-file and evaluates them.
It can optionally optimize the algorithms' numeric hyperparameters. Librec-Auto, however, does not return a trained model nor does it perform algorithm selection. It is an evaluation tool for a set of predefined algorithms.  
% Auto-CaseRec
Auto-CaseRec is an early prototype of an AutoRecSys framework. It solves the CASH problems the Case Recommender library \citep{daCosta:2018:CRF:3240323.3241611}. 
Yet, it does not offer the features and thorough development of a normal automated library. Mainly, its file-based usage and missing installation support (like a requirements file) are problematic. 
As a result, it was not usable as a part of our comparison without first reimplementing some of its functionalities. 
Consequentily, Auto-CaseRec and Librec-Auto are not sophisticated enough to be AutoRecSys. 

%% Discuss used Algorithms 
All algorithms in a library are integrated into our comparison if possible. 
We have used $10$ algorithms from LensKit, $11$ from Surprise, $26$ from scikit-learn, and $1$ algorithm from each other library. 
Details on which algorithms are used per library are listed in the Appendix \ref{apdx:usedAlgorithms}. 
Additionally, a baseline $B_{mean}$ was added. $B_{mean}$ always predicts the mean of the ratings calculated using the training data. $B_{mean}$ is part of its own category "Baseline".
In total, $60$ algorithms are compared. 

% 10 (Lenskit) + 11 (Surprise) + 1 (Spotlight) + 1 (Xgboost) + 26 (Sklearn) + 1 (ktrain) + 1 (baseline) + 1 (AutoRec) + 1 (Auto-Surprise) + 7 (All AutoML Libs) 
\subsection{Experiment Setup}
%Our experiment was setup using the following steps. 
\subsubsection*{Preprocessing}
%% Discuss the difference between RecSys and ML
We preprocessed the RecSys datasets such they are usable by (Auto)RecSys and (Auto)ML algorithms and as close as possible to the original dataset. Details on the chosen preprocessing steps can be found in the Appendix \ref{apdx:preprocessing}.
To present an overview of the resulting datasets, see Table \ref{tab:datastatistics} for details on the datasets' statistics. 
Table \ref{tab:datastatistics} shows the number of customers, items, and instances of each preprocessed dataset. It also shows the number of additional features. Additional features are all features excluding customer-, item-ID and timestamp. % and rating. 
Furthermore, the table details how many ratings each customer has done on average and the minimum/maximum number of ratings by a single customer.
\begin{table}
    \centering
    \caption{\textbf{Dataset Statistics}
    For each dataset, the number of customers, items, instances, and additional features are shown. Additional features represent the number of features besides customer-, item-ID, and timestamp. Statistics about the Ratings per Customer (RpC) are also shown.}
    \label{tab:datastatistics}
    \resizebox{\textwidth}{!}{\begin{tabular}{lS[scientific-notation=true] S[scientific-notation=true] S[scientific-notation=true] S[table-format=2.0, round-precision=0] S[table-format=3.2, round-precision=2] S[table-format=2.0, round-precision=0] S[table-format=4.0, round-precision=0] S[table-format=2.0, round-precision=0]}
        \toprule
        \text{Dataset} & \text{\#customers} & \text{\#items} & \text{\#instances} & \text{\#$add_{features}$} & \text{Avg. RpC} & \text{Min. RpC} & \text{Max. RpC}\\
        \midrule
        amazon-appliances & 47 & 47 & 190 & 4 & 4.042553191489362 & 1 & 8 \\
        amazon-digital-music & 1485 & 185 & 2243 & 4 & 1.5104377104377105 & 1 & 14 \\
        amazon-fashion & 406 & 31 & 3076 & 4 & 7.5763546798029555 & 1 & 16 \\
        amazon-software & 1826 & 802 & 11981 & 4 & 6.561336254107339 & 1 & 52 \\
        amazon-industrial-and-scientific & 11041 & 5327 & 72431 & 4 & 6.560184765872656 & 1 & 92 \\
        movielens-100K & 943 & 1682 & 100000 & 44 & 106.04453870625663 & 20 & 737 \\ 
        movielens-latest-small & 610 & 9724 & 100836 & 20 & 165.30491803278687 & 20 & 2698 \\
        movielens-1M & 6039 & 3705 & 998080 & 22 & 165.27239609206822 & 19 & 2314 \\
        foodCom & 226570 & 231637 & 1132367 & 12 & 4.997868208500684 & 1 & 7671 \\
        amazon-toys-and-games & 208180 & 78698 & 1759314 & 4 & 8.450927082332596 & 1 & 718 \\
        amazon-movies-and-tv & 297529 & 60110 & 3293658 & 4 & 11.070040231372404 & 1 & 3398 \\
        amazon-electronics & 728719 & 159935 & 6537585 & 4 & 8.971338746485271 & 1 & 576 \\
        yelp & 2189457 & 160585 & 8635384 & 58 & 3.9440756315378653 & 1 & 6073 \\
        netflix & 458382 & 17759 & 10000000 & 1 & 21.81586536993163 & 1 & 1779 \\
        \bottomrule
    \end{tabular}}
\end{table}

\subsubsection*{Hyperparameter Settings}
\label{sec:hyperparameterSettings}
For all algorithms, the default values for hyperparameters are used. 
% No default paras
If an implementation does not provide default parameters, the parameter values of the sample code from the implementation's documentation were used, e.g., values in the getting started section.

% Automated libraries resource default 
For automated libraries, resource parameters were set to have the same memory, job count, time passed to the framework (called search time), and optimization metrics across all libraries. 
We set the memory to all available memory (528 GB), the number of jobs to the maximal number of cores (64 - one for each CPU thread), the search time to $4$ hours elapsed real (wall-clock) time, and the metric to the RMSE.
% Special cases for automated libraries 
Not all automated libraries support the RMSE. In such cases, the library was adjusted to optimize the MSE. 
Furthermore, not all libraries provide parameters to control memory or job count. In these cases, the behavior of the library is to use all memory or cores.

% Iterative Algorithm Settings
RecSys and ML libraries do not necessarily allow a user to set a time limit or tolerance for stopping the training of an algorithm.
Instead, libraries rely on a user-controlled parameter to define the maximum number of trials or epochs for training.
The default training parameters are not necessarily representative for the given algorithm while being faithful to our use case of an inexperienced user. 
To include an alternative in our analysis, we made an exception for AutoRec and set it to use $75$ search trials and $1000$ epochs with early stopping for all datasets.
% talk about the drawbacks from the above decision? 
% Choosing to only use default parameters makes the training of many libraries subpar. Specifically, libraries based on neural networks usually have a small number of training epochs by default. Nevertheless, even with the default values some algorithms run into the time limit for some datasets (e.g.: spotlight on netflix). 
% For us to use such libraries otherwise would require intensive re-implementation to make it usable with a time limit. Something the libraries should be able to offer themselves. 

\subsubsection*{Resource Limits}
\label{sec:ResourceLimits}
We implemented a runtime limit. 
If an algorithm takes more than $4$ hours and $30$ minutes for its evaluation, its execution is terminated. 
For automated libraries, this runtime limit is extended by the search time. Hence, totaling a limit of $8$ hours and $30$ minutes. 
An algorithm is also terminated if its evaluation exceeds $528$ GB, the maximal available memory of our system. 

\subsubsection*{Predictive Performance Evaluation}
To compare the predictive accuracy across datasets, we transformed the RMSE into a rank. 
A lower rank is better (e.g.: 1 is the best rank). 
All algorithms that ran into a limit were ranked last. Equally well-performing algorithms are assigned the average of their combined ranks.

\section{Results}
\label{sec:results}
% Objectively describe your results
We first present the impact of our chosen resource limits. Next, we evaluate the predictive performance of the algorithms and their associated categories.

\subsection{Impacts of Resource Limits}
\label{sec:impactsOfLimits}
%% 1. Talk about how many unsuccessful algorithms on which datasets per category 
Each evaluation run was limited in terms of memory and time. 
$7.50\%$ of all evaluation runs failed. The memory limit was reached in $3.81\%$ and the time limit in $3.69\%$ of all runs. This, however, depends very much on the dataset. See Table \ref{tab:failureDatasets} for an overview over the different datasets. 
\begin{table}
    \centering
    \caption{\textbf{Failure of Algorithms per Category per Datasets}
    \normalfont Shown are the absolute failure counts and the failure frequencies for each dataset per category. The number next to the category name shows the number of algorithms in this category.
    The baseline $B_{mean}$ is not shown, it did not run into any limits. However, it is part of the total count of algorithms for the last column. 
    }
    \label{tab:failureDatasets}
    \resizebox{\textwidth}{!}{\begin{tabular}{lcccc|c}
        \toprule
        Dataset & AutoML (7) & AutoRecSys (1) & ML (28) & RecSys (23) & Total (60) \\
        \midrule
        amazon-appliances&-&-&-&-&-\\
        amazon-digital-music&-&-&-&-&-\\
        amazon-fashion&-&-&-&-&-\\
        amazon-software&-&-&-&-&-\\
        amazon-industrial-and-scientific&-&-&-&-&-\\
        movielens-100K&-&-&1 (3.57\%)&-&1 (1.67\%)\\
        movielens-latest-small&-&-&1 (3.57\%)&-&1 (1.67\%)\\
        movielens-1M&-&-&3 (10.71\%)&-&3 (5.00\%)\\
        foodCom&-&-&3 (10.71\%)&2 (8.70\%)&5 (8.33\%)\\
        amazon-toys-and-games&1 (14.29\%)&-&3 (10.71\%)&5 (21.74\%)&9 (15.00\%)\\
        amazon-movies-and-tv&-&-&3 (10.71\%)&5 (21.74\%)&8 (13.33\%)\\
        amazon-electronics&-&-&3 (10.71\%)&6 (26.09\%)&9 (15.00\%)\\
        yelp&4 (57.14\%)&-&4 (14.29\%)&8 (34.78\%)&16 (26.67\%)\\
        netflix&1 (14.29\%)&-&4 (14.29\%)&6 (26.09\%)&11 (18.33\%)\\
        \bottomrule
    \end{tabular}}
\end{table}

%% 2. Talk about yelp failure for automl tools
The majority of AutoML tools failed on the yelp dataset. The yelp dataset requires at least ${\sim}1.6$ GB memory. This was too much for AutoPyTorch, AutoSklearn, and TPOT. Both AutoPytorch and AutoSklearn limit the memory per job. For the yelp dataset, the memory limit per job is quickly exceeded if the memory (528 GB) is split between all 64 jobs. TPOT also runs in a job-related memory limit.  
All other AutoML failures were timeouts. 
In these cases, GAMA, TPOT, and AutoGluon were not able to build a model from the training data and generate the final predictions for the test data in the provided $4$ hours and $30$ minutes after the $4$-hour-long search. 
 
%% 3. Talk about algorithms randomly (however biased as a lot of "bad scaling" algorithms are included by basic ML/RecSys algorithms)
Assuming an inexperienced user would randomly pick an algorithm to solve a dataset, the algorithm would not produce any result within our resource limitations in $7.50\%$ of all cases. For yelp, this would occur with a probability of $26.67\%$. 

%% 4. AutoRec and failure vs. other deep learning appraoches with default values 
Recall, we made an exception for AutoRec, a RecSys library, and let it run for more epochs and trials than the default values. This made AutoRec run in a timeout for all six datasets with more than a million instances. 
Another deep learning RecSys library, Spotlight, which only runs for ten iterations by default, did run in a timeout for netflix and yelp.
Yet, in all eight cases where both did not run into a limit, AutoRec and Spotlight are tied for the number of better predictive accuracy scores. In other words, neither default values nor static custom values are dominant. 

\subsection{Predictive Performance}
To inspect the categories' performance, see Table \ref{tab:avgRankCat} for the categories' average rank. This data indicates the category ranking: $AutoRecSys\:(3.57) > AutoML\:(18.64) > RecSys\:(28.03) > ML\:(36.20) > Baseline\:(37.64)$. Since the AutoRecSys category contains only one library (AutoSurprise), the average of the AutoRecSys category equals the result of AutoSurprise. AutoSurprise did run on all datasets and did neither experience memory limits nor timeouts. 
Next, we switch to an algorithm-specific perspective. % to lessen the impact of the average.

Refer to Table \ref{tab:top10ranked} for an overview of the top 10 ranked algorithms. 
The top 10 includes all categories except for the baseline. The AutoRecSys library Auto-Surprise allocates the first place with a substantial average rank difference to the second place. Auto-Surprise is followed by the two AutoML tools FLAML and H2O. The majority of algorithms within the top 10 are from RecSys libraries. One ML algorithm, XGBoost, reached the top 10.
FLAML's average rank is exactly twice the average rank of Auto-Surprise. In contrast, the rest of the average ranks differ only marginally between subsequent placings. 
\begin{table}
    \parbox{.45\linewidth}{
        \centering
        \captionsetup{margin=0.05cm}
        \caption{\textbf{Average Rank per Category} 
        %The table shows the average rank of a category and the failure percentage of algorithms in this category over all datasets. 
        The number next to the category shows the amount of algorithms in the category.
        }
        \label{tab:avgRankCat}
        \begin{tabular}{l S[table-format=2.3, round-precision=2] c}
            \toprule
            Category & {Average rank} & {Failure $\%$} \\
            \midrule
            AutoRecSys (1)&3.5714285714285716&-\\
            AutoML (7)&18.642857142857142& 6.12\%\\
            RecSys (23)&28.032608695652176&9.94\%\\
            ML (28)&36.19770408163265& 6.38\%\\
            Baseline (1)&37.642857142857146&-\\
            \bottomrule
        \end{tabular}
    }
    \hfill
    \parbox{.45\linewidth}{
        \centering
        \caption{\textbf{Top 10 Average Ranked Algorithms}}
        \label{tab:top10ranked}
        \begin{tabular}{lS[table-format=2.3, round-precision=3]S[table-format=1.4, round-precision=4]}
            \toprule
            Algorithm & \text{Average rank}\\
            \midrule
            Auto-Surprise&3.5714285714285716\\
            FLAML&7.142857142857143\\
            H2OAutoML&7.214285714285714\\
            SVDpp (Surprise)&8.071428571428571\\
            FunkSVD (LensKit)&9.357142857142858\\
            BiasedSVD (LensKit)&11.178571428571429\\
            SVD (Surprise)&11.285714285714286\\
            BaselineOnly (Surprise)&13.071428571428571\\
            ALSBiasedMF (LensKit)&13.571428571428571\\
            XGBoostRegressor&15.5\\
            \bottomrule
        \end{tabular}
    }
\end{table}

\begin{figure}
  \centering
  \includegraphics[width=\textwidth]{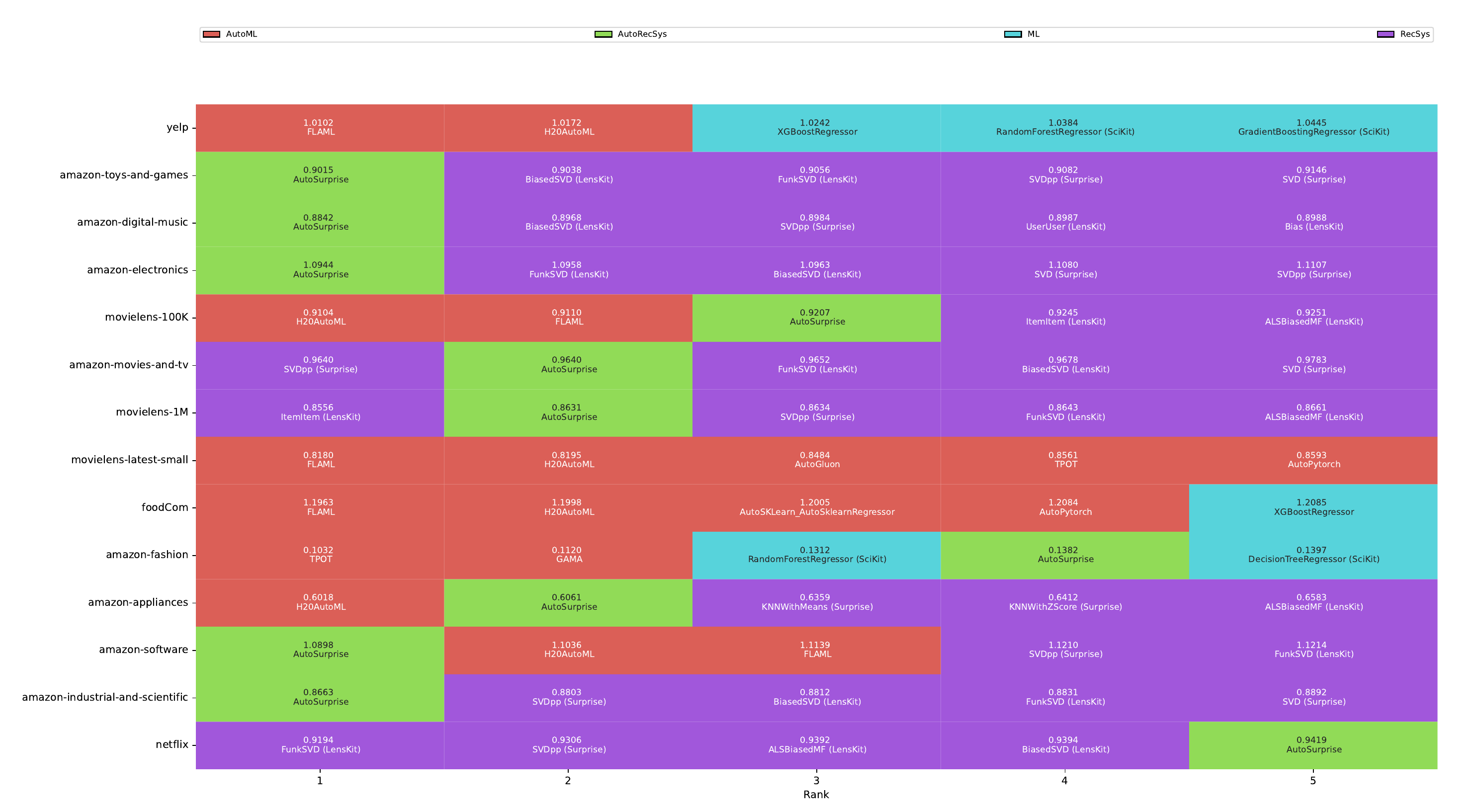}
   %\Description{Several algorithms per dataset are shown. RecSys algorithms are most dominant on average. Auto-Surprise and AutoML algorithm are most often in first place. For the yelp dataset, ML algorithm are dominant.}
  \caption{\textbf{Top 5 Algorithms per Dataset}
            For each dataset, the top 5 algorithms are shown. They are color coded for their category. Moreover, the RMSE and name of the algorithm are shown. A larger, more readable version of this figure is located in the appendix.
            }
  \label{fig:top5AlgoPerDataset}
\end{figure}

%% Rank plot for top 5 
For more detailed ranking information per dataset, see Figure \ref{fig:top5AlgoPerDataset} for a rank plot of algorithms per dataset.
% - 6 out of 8 first places are Automated Libarys
The datasets' top 5 are dominated by RecSys algorithms, which are most often a variation of SVD. 35 out of 70 are RecSys algorithms and 11 out of 70 are AutoRecSys. In total, 46 out of 70 are RecSys or AutoRecSys (65.7\%). 
Yet, RecSys algorithms are only able to secure the first place three times. The first place is dominated by %  the automated libraries with 
AutoRecSys (5 times) and AutoML (6 times). 

% - large datasets that include many add_features (yelp) is dominated by ML librarys
Datasets with a lot of additional features, like yelp or movielens-latest-small, are mostly dominated by ML approaches except for movielens-1M and movielens-100k. 
Movielens-1M and movielens-100k are extremely popular evaluation datasets in the RecSys community \citep{beel2019data} and thus a lot of tools are tuned to work best on these datasets by default. 
Similar, datasets with almost no additional features like netflix are dominated by RecSys algorithms. For datasets with some additional features, the results are mixed. 
% - marginal difference in error 
%The rank differences are sometimes caused only by a marginal difference in the RMSE. For example, the first and second place for amazon-movies-and-tv have the same RMSE. In contrast, larger differences can also exist, like the difference between the first and second place for netflix. 

\begin{figure}
  \centering
  \includegraphics[width=\textwidth]{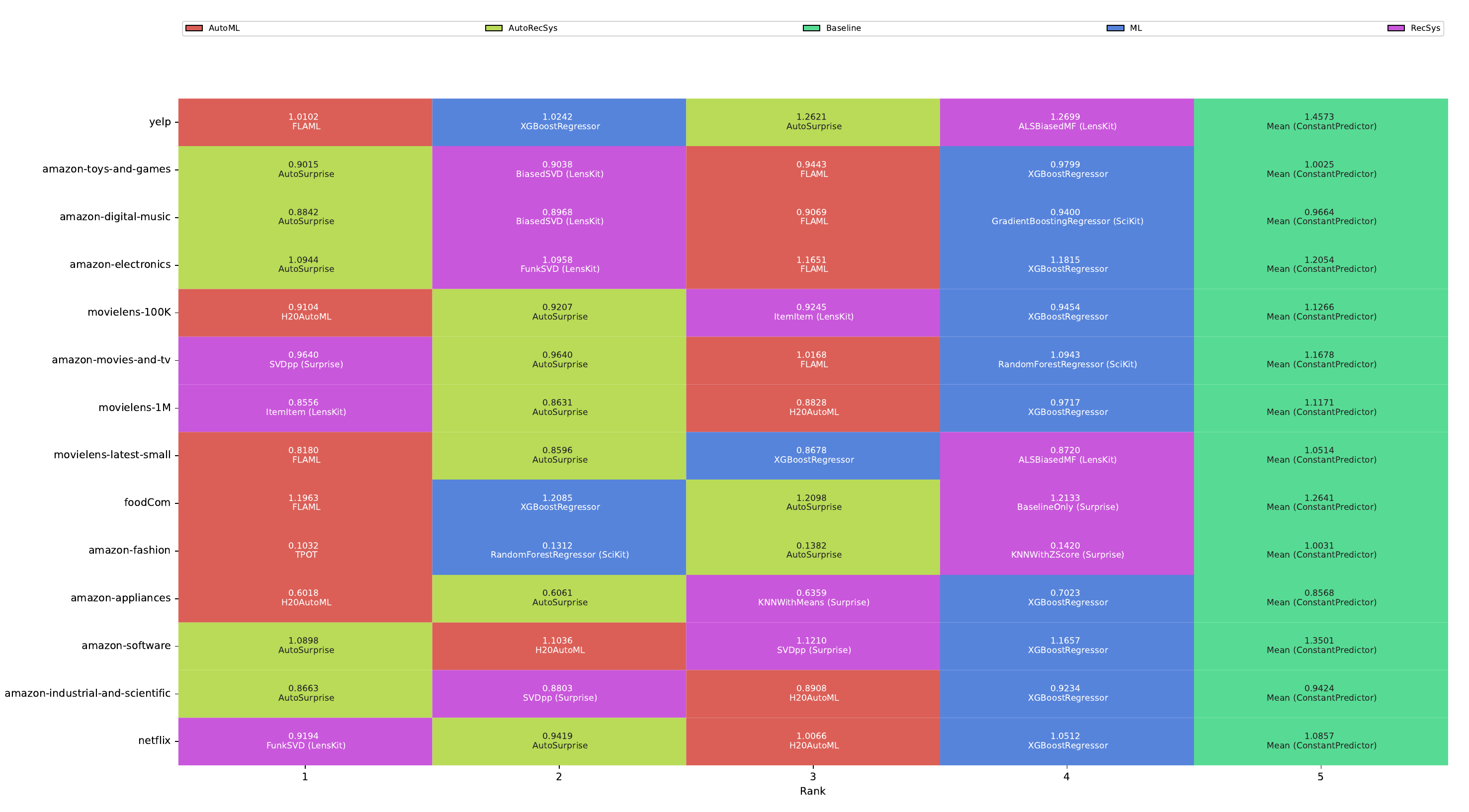}
  \caption{\textbf{Order of Categories per Dataset}
            For each dataset, the 5 categories are shown sorted by their best performing algorithm's RMSE. They are color coded for their category. Moreover, the RMSE and name of the best performing algorithm are shown. A larger, more readable version of this figure is located in the appendix.
            }
  %\Description{The mean constant predictor baseline is always in the last place. The first 3 places are split between automl, autorecsy and recsys. ML is usually in the foruth place. The first place is allocated most often by automl or autorecsys.}
    \label{fig:top5CatPerDataset}
\end{figure}

Besides an algorithm rank perspective, a category rank perspective is relevant. See Figure \ref{fig:top5CatPerDataset} for a rank plot of categories per dataset. 
Our chosen baseline is, as expected, always the worst category. This is closely followed by ML, which is in the fourth place 10 out of 14 times. Automated libraries are never in the fourth place. 
The first and second place are dominated by AutoRecSys and AutoML, with AutoRecSys being in the first or second place 11 times and AutoML 7 times. 
This follows the category ranking indicated by the tables above, while the first place is more disputed. 


%% 2.2 Present something with the boxplots of categories (Too large errors and that we filter them here)
\paragraph{\textbf{Comparison to the Baseline}} Many outliers exist in the collected error values\footnote{See Appendix \ref{apdx:boxplot} for more details.}. 
Such outliers are expected since all algorithms with default values within a library were used.
% 226 form 840 (26.90%) algorithms have a RMSE worse than the baseline.
However, it is unexpected that the baseline $B_{mean}$, which only predicts the mean, outperforms other algorithms frequently. On average across all datasets, $B_{mean}$ outperforms $26.90\%$ of all algorithms. 

\section{Discussion}
%% Specifically answer research question 
Given our results, we can answer the research question: 
\emph{What is the performance of AutoML algorithms on RecSys tasks compared to AutoRecSys, ML, and RecSys algorithms using default hyperparameters?}
The results indicate the following performance ranking for default hyperparameters: $AutoRecSys \geq AutoML > RecSys > ML > Baseline$ whereby the first place is disputed between AutoRecSys and AutoML. 
In other words, we recommend AutoML or AutoRecSys libraries to an inexperienced user.
AutoML is a promising approach to RecSys with a lot of potential considering that AutoML is not (over)fitted to RecSys tasks and much more accessible than RecSys libraries. 

% RecSys vs. ML and its consequences 
Given that RecSys algorithms often outperform ML algorithms but AutoML still outperforms RecSys, the need for more sophisticated AutoRecSys libraries is shown. Likewise, extensions of AutoML frameworks to RecSys have a lot of potential, given that AutoML is able to outperform the application-specific AutoRecSys or RecSys in $6$ of $14$ RecSys datasets. 
Moreover, the accessibility of RecSys libraries is in stark contrast to the accessibility of AutoRecSys or AutoML libraries (see our code for examples). 

% Failure 
In general, the percentage of failed runs and bad performances is surprisingly high. 
Assuming an inexperienced user would randomly pick an algorithm, the algorithm would not produce any result within our resource limits or perform worse than the simple baseline $B_{mean}$ in $34.40\%$ of all cases.
While this perspective is biased by bad performing algorithms and our resource limits, it is still truthful to our inexperienced-user-scenario. An inexperienced user does not know which algorithms are inappropriate or badly implemented. Furthermore, an inexperienced user has almost no guidelines about which algorithm to use in RecSys. 
%To avoid worse than baseline performance, the algorithms would need to be drastically filtered. 
Likewise, the impact of hyperparameter values on the performance and failure, like the number of training iterations for AutoRec and Spotlight, further illustrates the need for automated libraries to make RecSys more accessible.

% Benchmarks in RecSys
Using benchmarking suites is a standard practice in AutoML. In RecSys, methods are evaluated on a few (potentially not diverse) datasets. However, we can see that a difference in performance of methods between such common datasets (like movielens-1M, movielens-100k, netflix) and more uncommon datasets (like Food.com and Yelp with our preprocessing) exist.
The need for a diverse and complete benchmarking suite for RecSys is apparent. 

\section{Conclusion}
% The 9 pages allocated for the main paper must include a discussion of limitations 
% and a broader impact statement regarding the approach, datasets and applications 
% proposed/used in your paper. It should reflect on the environmental, ethical and 
% societal implications of your work. The statement should require at most one page.
%
% This section is included in the template as a default, but you can also place these
% discussions anywhere else in the main paper, e.g., in the introduction/future work.
To conclude our work, we first recap and discuss the broader impact of our work. Then, we detail the potential directions of improvement. Next, the limitations of our work are explicitly stated. Finally, the future work is addressed. 

\subsubsection*{Summary and Broader Impact Statement}
\label{sec:broaderImpactStatement}
We compared the predictive performance of $60$ AutoML, AutoRecSys, ML, and RecSys algorithms, including a baseline, on $14$ RecSys datasets with explicit feedback.
The open-source implementation of our comparison tool allows others to evaluate algorithms' performance on RecSys datasets. 
We presented algorithm performance through a ranking calculated using RMSEs obtained by hold-out validation. 
The presented results allow users to compare algorithms for a real-world RecSys application. 
Specifically, the results indicated $AutoRecSys \geq AutoML > RecSys > ML > Baseline$ as a performance ranking. Moreover, we have shown that it is likely to encounter worse-than-baseline performing algorithms.   

%% Lead to automated approaches + pros and cons  
A positive impact of the presented results is that users are able to select an appropriate algorithm or category of algorithms without being RecSys experts. 
The results could lead more users to AutoML or AutoRecSys.
AutoML or AutoRecSys can help users to create real-world RecSys applications without deep RecSys knowledge. Yet, if users do not know the risks of RecSys, their models could lead to biased, unfair, or offensive recommendations. Such misuse could create filter bubbles and bubbles that support malicious content. 

% other positive impacts 
Generally, our work is a first step towards fair and comparable benchmarking in RecSys. 
% (What work could be done to increase the positive impacts and reduce the downsides?)
Initiatives to make data more standardized and easily shareable would be highly impactful for the RecSys community. Likewise, extending AutoML libraries to RecSys applications can create a positive impact. Alternatively, an initiative to support and produce sophisticated AutoRecSys libraries would help as well. 
The downsides of comparing different algorithms and making it easier to select appropriate algorithms, like automated approaches, could be reduced by increasing the awareness about the risks of automated approaches in RecSys. 

\subsubsection*{Limitations}
\label{sec:limitations}
% - not implicit feedback, a protocol to use ML tools for implicit 
Our comparison focused on RecSys datasets with explicit feedback. Therefore, we did not include datasets with implicit feedback. 
% - no NLP for features, only basic preprocessing, a lot of data-specific improvements 
The preprocessed datasets used for evaluation are basic by design and without natural language processing. 
% - no benchmarking suite due to licenses
These preprocessed datasets are not redistributed. Only scripts to prepossess the original datasets are provided.  
% - features used by algorithms 
The selected RecSys algorithms can only use a user-ID, item-ID, rating, and optionally timestamps from a dataset.  
We did not constrain (Auto)ML algorithms to use a subset of the available features. As a result, RecSys algorithms use fewer features than ML algorithms.
% - resource limits
We worked with resource limitations and not all evaluations were able to finish within our limits.
%  - too few datasets, libraries 
Additional datasets, libraries, and more expensive validation methods (like cross-validation) are missing in this first version of our comparison.

\subsubsection*{Directions of Improvement}
Our results hint at the effectiveness of automated libraries for inexperienced users in RecSys. 
Moreover, we have only found one suitable automated library that explicitly focused on RecSys.
Consequently, there is a clear need for better and more sophisticated automated libraries for RecSys.
We think that the sophisticated AutoML libraries need to get extensions for RecSys or better AutoRecSys libraries need to be made. 
Improvements in these directions shall end the manual and not systematically automated search for suitable configurations in RecSys.
Thus, bringing the advances from the AutoML community to the RecSys community. Such that, in turn, the RecSys community will focus more on AutoML in the future.

\subsubsection*{Future Work}
%% Future Work 
%   - For implicit feedback
%   - with NLP processing ("the impact of NLP on recsys")
%   - more datasets, libraries and validations 
% automl extension (to use implict feedback or have specific protocols for recsys)
In future work, a better comparison between AutoML, AutoRecSys, ML, and RecSys algorithms can be achieved. 
% Cross validation 
Firstly, extending our comparison to perform cross-validation with a bigger resource budget is necessary. Thus, reducing the noise of randomness and to get more meaningful results. 
% more datasets, more algorithms
Secondly, the comparison needs to include more datasets and algorithms.
% implicit feedback 
Lastly, the comparison of algorithms and categories must be extended to implicit feedback datasets. 
%In the last couple of years, the RecSys community focused more on implicit than explicit feedback. 
%Therefore, it is important to include implicit feedback in RecSys benchmarks.
%% Be better. 
Besides achieving a better comparison, a detailed analysis of such a comparison is future work. 
Determining the reason for the difference in the performance of compared algorithms can help to improve existing algorithms or foster the creation of new algorithms.

% print bibliography -- for bibtex / natbib, use:
\bibliography{references}

% and for biber / biblatex, use:

%\printbibliography

% supplemental material -- everything hereafter will be suppressed during
% submission time if the hidesupplement option is provided!

\newpage
\appendix

\section{Versions, URLs, and Licenses of all used Libraries and Datasets}
\label{apdx:lib_data}
\subsection{Libraries}
Surprise, \rurl{github.com/NicolasHug/Surprise}, 1.1.1, BSD 3-Clause;
LensKit, \rurl{lenskit.org}, 0.13.1, Custom;
Spotlight, \rurl{github.com/maciejkula/spotlight}, 0.1.6, MIT;
AutoRec, \rurl{github.com/datamllab/AutoRec}, 0.0.2, None;
Auto-Surprise, \rurl{github.com/BeelGroup/Auto-Surprise}, 0.1.7, MIT;
scikit-learn,  \rurl{scikit-learn.org}, 1.0.1, BSD 3-Clause;
XGBoost, \rurl{github.com/dmlc/xgboost}, 1.5.1, Apache-2.0;
ktrain, \rurl{github.com/amaiya/ktrain}, 0.28.3, Apache-2.0;
Auto-sklearn, \rurl{automl.github.io/auto-sklearn}, 0.14.2, BSD 3-Clause; 
FLAML, \rurl{github.com/microsoft/FLAML}, 0.9.1, MIT;
GAMA, \rurl{github.com/openml-labs/gama}, 21.0.1, Apache-2.0;
H2O, \rurl{h2o.ai/products/h2o-automl}, 3.34.0.3, Apache-2.0; 
TPOT, \rurl{github.com/EpistasisLab/tpot}, 0.11.7, LGPL-3.0;
AutoGluon, \rurl{auto.gluon.ai/stable/index.html}, 0.3.1, Apache-2.0;
Auto-PyTorch, \rurl{github.com/automl/Auto-PyTorch}, 0.1.1, Apache-2.0;

\subsection{Datasets}
MovieLens 100k, \rurl{grouplens.org/datasets/movielens/100k/}, Custom;
MovieLens 1M \rurl{grouplens.org/datasets/movielens/1m/}, Custom;
MovieLens Latest 100k (9/2018), \rurl{grouplens.org/datasets/movielens/latest/}, Custom;
Amazon Review Data, \rurl{nijianmo.github.io/amazon/index.html}, N/A;
Yelp Open Dataset, \rurl{yelp.com/dataset}, Custom;
Netflix Prize Dataset, \rurl{www.kaggle.com/netflix-inc/netflix-prize-data}, Custom; 
Food.com Recipes and Interactions Dataset, \rurl{https://www.kaggle.com/shuyangli94/food-com-recipes-and-user-interactions}, Data files © Original Authors (ODC-ODbL);

\section{Used Algorithms}
\label{apdx:usedAlgorithms}
The following algorithms of different libraries were used:
\begin{description}
\item[Surprise:] SVD, KNNBasic, KNNBaseline, KNNWithZScore, KNNWithMeans, CoClustering, BaselineOnly, SlopeOne, SVDpp, NMF, NormalPredictor
\item[LensKit:] ItemItem, UserUser, Bias, ALSBiasedMF, FunkSVD, BiasedSVD, TFBiasedMF, BPR, IntegratedBiasMF, HPF
\item[Sklearn:] LinearRegression, Ridge, SGDRegressor, ElasticNet, LassoLars, OrthogonalMatchingPursuit, ARDRegression, BayesianRidge, Lars, Lasso, HuberRegressor, TheilSenRegressor, PoissonRegressor, GammaRegressor, TweedieRegressor, RANSACRegressor, RandomForestRegressor, AdaBoostRegressor, BaggingRegressor, ExtraTreesRegressor, GradientBoostingRegressor, SVR, KNeighborsRegressor, MLPRegressor, KernelRidge, DecisionTreeRegressor
\end{description}
All other libraries only have one relevant implementation for regression.

\section{Preprocessing}
\label{apdx:preprocessing}
The goal of all preprocessing steps was to to keep the data as close to the original as possible. Only features that would cause errors were preprocessed. All datasets were converted into CSV files. Such a CSV file shall be usable by RecSys and ML libraries. To achieve this, several decisions on how to transform features had to be made. 

% NLP free data
In general, we differentiate between categorical and text features. Text features can contain any free-text input (e.g., text of a review), while categorical features are limited to a specific range of possible categories. For this paper, it was decided to not include any text features. In other words, Natural Language Processing was not used to transform text features into any type of usable tabular features. Categorical features have been One-Hot encoded. 

% Transformation only if needed 
Based on the assumption to keep the data as close to the original as possible, an already encoded feature was used and not reencoded. For example, if the gender of a user was encoded through a dummy variable, it was not transformed to one-hot encoding. 
This also applies to customer- and item-IDs. If the ID was given as an integer, it was used as an integer. Only if the dataset used a different format, like a UUID in the form of a string which would prompt errors in ML or RecSys algorithms, the IDs were encoded as increasing integers. 
Any other numeric feature is used without processing. 

% no own feature creation
Similar to the previous assumption, we decided against creating new features from existing data to stick as close to the original data as possible. To illustrate, a feature representing URLs related to an item was removed instead of using these URLs to potentially fetch additional information about an item. 

% undocumented features 
Sometimes features were in the dataset but not in the dataset's documentation. In such cases, the undocumented features were removed. 

% Duplicate Handling
Removing features, like a review's text, can cause two reviews of an individual customer to become identical.
Moreover, some datasets have multiple entries for the same review if a property of the review differs. For example, in the Amazon datasets the "summary" of a review can be different and thus result in two distinct entries. 
Likewise, (presumably) bugs can cause that the timestamp of a review and the timestamp of an edited version are the same. In the Yelp dataset, multiple duplicates exist, because all features (including the timestamp) are identical but the text differs.  
We drop duplicates such that the combination of resulting features from previous preprocessing steps is unique.

% zero ratings 
Several algorithms can not handle ratings to be zero. For these specific cases, we add a small $\epsilon$ to all zero-ratings such that the algorithm is still usable. 

% we are using all files 
The defined preprocessing steps were applied to all features. If features were spread across multiple files, all files are used.  

\section{Dataset Boxplot}
\label{apdx:boxplot}
\begin{figure}
  \centering
  \includegraphics[width=\linewidth]{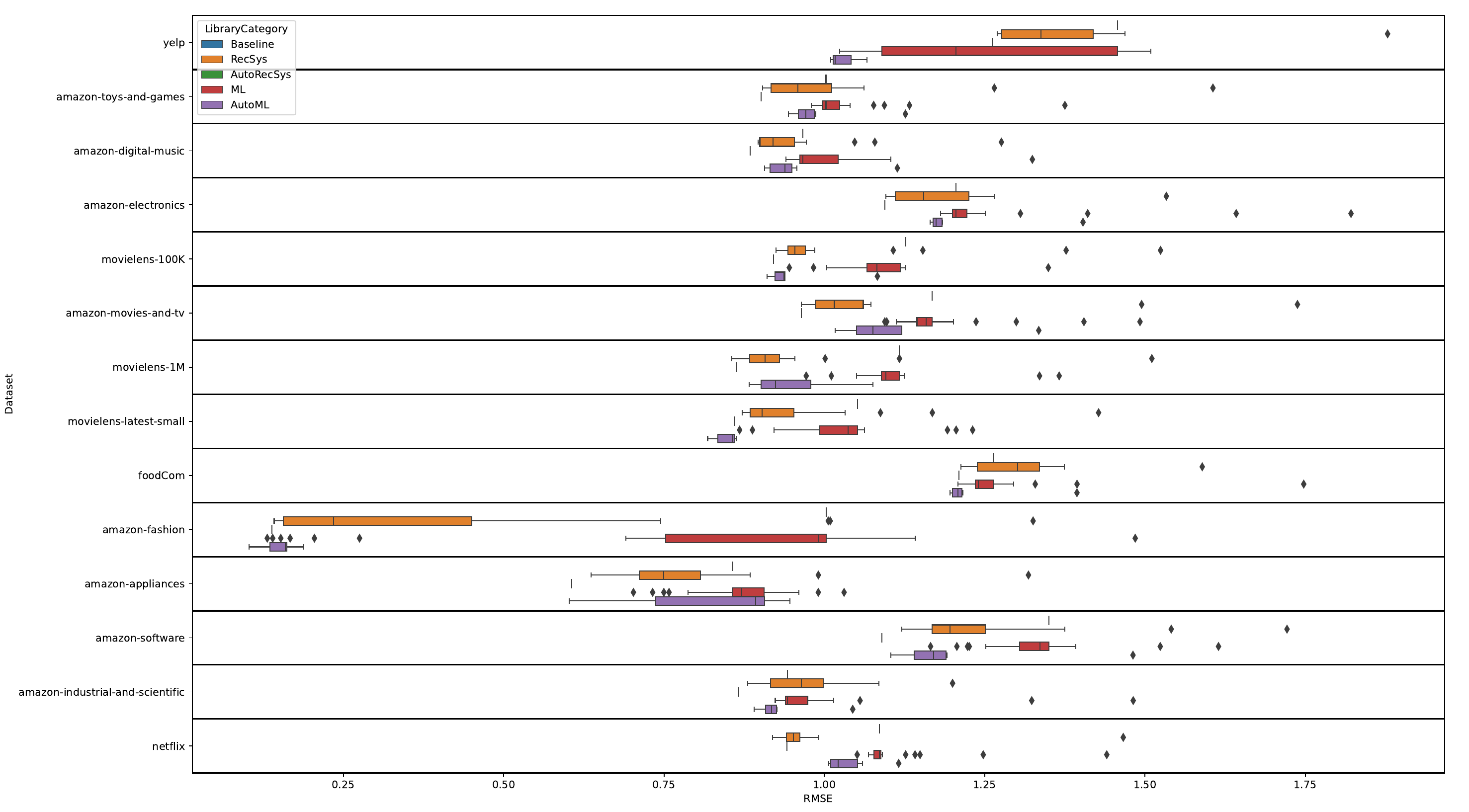}
  \caption{\textbf{RMSE Boxplots for each category specific to each dataset}
            The distribution of error values for each category specific to each dataset is visualized. The distribution is represented by boxplots. This excludes 61 error values that are greater than 2.
            }
  %\Description{Different depending on each dataset. Generally mixed results.}
    \label{fig:boxplotPerDataset}
\end{figure}

%% takaway from boxplot figure
% - amazon fashion low RMSE (very bad performance ML)
% - foodCom high RMSE
% - ML and RecSys are wider spread than the Auto tools (because of number of algorithms)
% - very mixed results, heavily varying distributions per dataset, no clear winner, many outliers  
Figure \ref{fig:boxplotPerDataset} shows the boxplots of the error values of the used algorithms for each category specific to each dataset.
The error values of most algorithms are between 0.8 and 1.3. Some error values for datasets stick out. The error values of the amazon-fashion datasets are comparatively low. The lowest values for the amazon-fashion dataset shown in figure \nameref{apdx:boxplot} are lower than 0.25. In comparison with the amazon-fashion- and other datasets, the error values of food.com are comparatively high.

\section{Larger Plots}

\begin{sidewaysfigure}
  \centering
  \includegraphics[width=\textwidth]{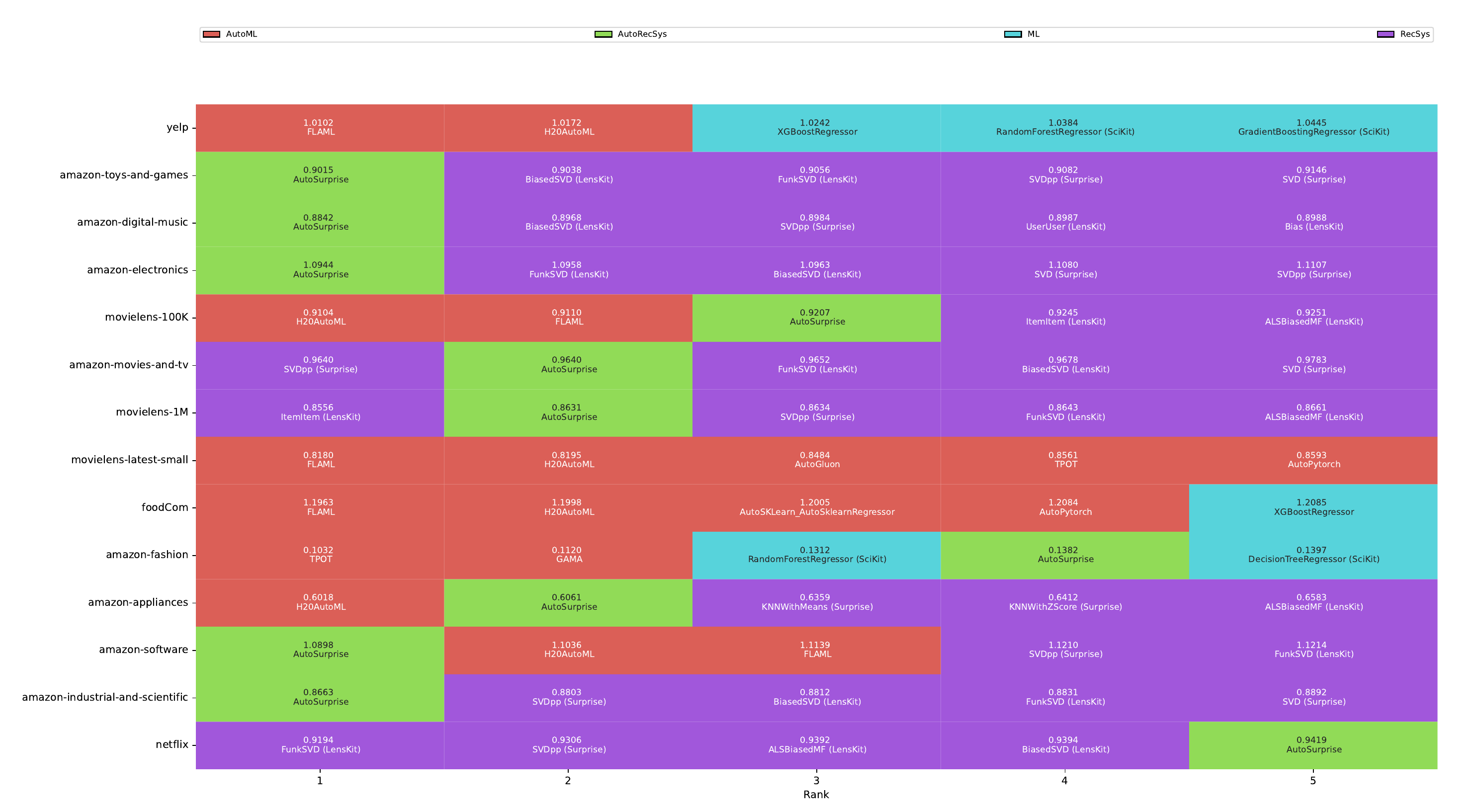}
  \caption{\textbf{Top 5 Algorithms per Dataset}
            For each dataset, the top 5 algorithms are shown. They are color coded for their category. Moreover, the RMSE and name of the algorithm are shown.
            }
\end{sidewaysfigure}
\begin{sidewaysfigure}
  \centering
  \includegraphics[width=\textwidth]{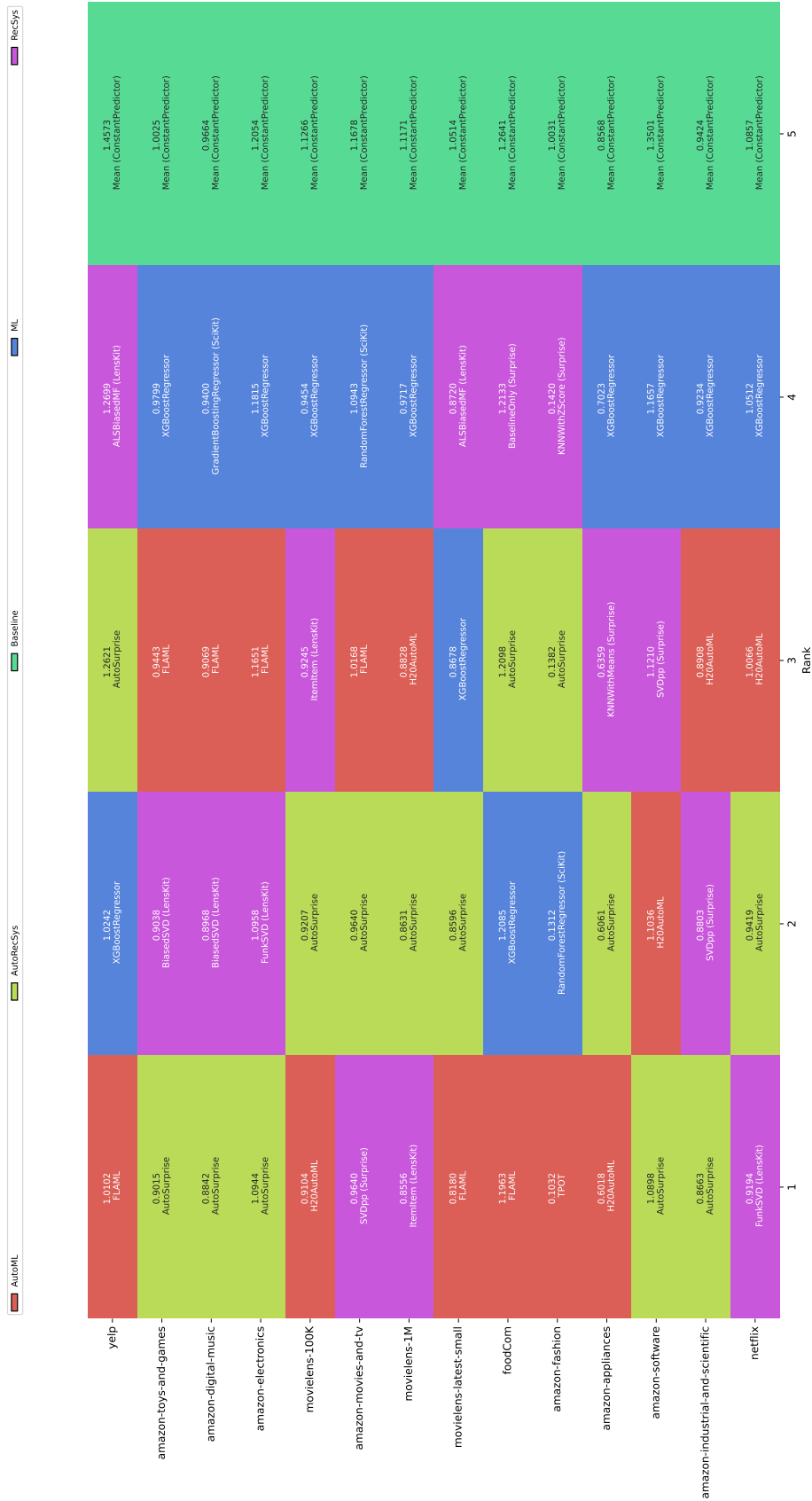}
  \caption{\textbf{Order of Categories per Dataset}
            For each dataset, the 5 categories are shown sorted by their best performing algorithm's RMSE. They are color coded for their category. Moreover, the RMSE and name of the best performing algorithm are shown.
            }
\end{sidewaysfigure}

\end{document}